\documentclass{aa}  
\usepackage{graphicx}
\usepackage{xcolor}
\usepackage{txfonts}
\usepackage{float}

\usepackage{hyperref}
\hypersetup{linkcolor=red,citecolor=blue,filecolor=cyan,urlcolor=magenta}
\begin{document} 
   \title{Statistical Study of Plasmoids associated with post-CME Current Sheet}

   \author{Ritesh Patel
          \inst{1,2}
          \and
            Vaibhav Pant \inst{3,4,5}
          \and
           Kalugodu Chandrashekhar \inst{6,7}
          \and
          Dipankar Banerjee \inst{1,2,8} }

   \institute{Indian Institute of Astrophysics, 2nd Block Koramangala, Bangalore 560034, India
         \and
         Aryabhatta Research Institute of Observational Sciences, Nainital 263001, India
        \and
            Instituto de Astrofis\'ica de Canarias, 38205 La Laguna, Tenerife, Spain 
        \and
            Departamento de Astrofis\'ica, Universidad de La Laguna, 38206 La Laguna, Tenerife, Spain\\
            \email{vaibhavpant55@gmail.com}
        \and
            Centre for mathematical Plasma Astrophysics, Department of Mathematics, KU Leuven, B-3001, Belgium
        \and
        Institute of Theoretical Astrophysics, University of Oslo, P.O. Box 1029 Blindern, N-0315 Oslo, Norway
        \and
        Rosseland Centre for Solar Physics,  University of Oslo, P.O. Box 1029 Blindern, N-0315 Oslo, Norway
        \and
            Center of Excellence in Space Science, IISER Kolkata, Kolkata 741246, India \\
             }

   \date{Received ; accepted Month DD, 2020}

 
  \abstract
{Coronal Mass Ejections (CMEs) are often observed to be accompanied with flare, current sheets, and plasmoids/plasma blobs. 2D and 3D numerical simulations and observations reported plasmoids moving upwards as well as downwards along the current sheet.}
{ We investigate the properties of plasmoids observed in the current sheet formed after an X-8.3 flare followed by a fast CME eruption on September 10, 2017 using {Extreme Ultraviolet} (EUV) and white-light coronagraph images. The main goal is to understand the evolution of plasmoids in different spatio-temporal scales using existing ground- and space-based instruments.}
{We identified the plasmoids manually and tracked along the current sheet in the successive images of {\it Atmospheric Imaging Assembly} (AIA) taken at 131 \AA\ pass-band and running difference images of white-light coronagraphs, K-Cor and LASCO/C2. The location and size of the plasmoids in each image is recorded and analysed covering the current sheet from the inner to outer corona.
}
{We found that the observed current sheet has an Alfv\'en Mach number of 0.018-0.35. 
The fast reconnection is also accompanied by plasmoids moving upwards and downwards. We identified 20 downward moving and 16 upward moving plasmoids using AIA 131 \AA\ images. In white-light coronagraph images only upward moving plasmoids are observed. Our analysis showed that the downward moving plasmoids have average width of 5.92 Mm whereas upward moving blobs have average size of 5.65 Mm in the AIA field of view (FOV). The upward moving plasmoids when observed in the white-light images have average width of 64 Mm in the K-Cor which evolves to a mean width of 510 Mm in the LASCO/C2 FOV. Upon tracking the plasmoids in successive images, we found that downward and upward moving plasmoids have average speeds of $\sim$272 km s$^{-1}$ and $\sim$191 km s$^{-1}$ respectively in the EUV channels of observation. The average speed of plasmoids increases to $\sim$671 km s$^{-1}$ and $\sim$1080 km s$^{-1}$ in the K-Cor and LASCO/C2 FOVs respectively implying that the plasmoids become super-Alfv\'enic when they propagate outwards. The downward moving plasmoids show an acceleration in the range of -11 km  s$^{-2}$ to over 8 km  s$^{-2}$. We also found that the null-point of the current sheet is located at $\approx$ 1.15 R$_\odot$ where bidirectional plasmoid motion is observed. }
{The width distribution of plasmoids formed during the reconnection process is governed by a power law with an index of -1.12.  Unlike previous studies there is no difference in trend for small as well as large scale plasmoids.
The evolution of width W of the plasmoids moving with average speed {\it V} along the current sheet is governed by an empirical relation, {\it V} = 115.69{\it W$^{0.37}$}. The presence of accelerating plasmoids near the neutral point indicates longer diffusion region as predicted by MHD models.}

\keywords{Sun: corona -- Sun: flares -- Magnetic reconnection -- Instabilities -- Methods: statistical
               }

   \maketitle
%

\section{Introduction}
Coronal Mass Ejections (CMEs) are the large expulsions of plasma and magnetic field in to the interplanetary space, observed as a distinct bright structure moving radially outwards in the white-light coronagraph images \citep{Hundhausen84}. Their origin have been primarily accounted by two mechanisms. One, from the prominence eruptions and the other, due to flaring active regions \citep{NatG2002}. Magnetic reconnection has been the driver for these flares and a number of models have been proposed to explain such phenomenon. The classical Sweet-Parker model \citep{Sweet, Parker} was the first theoretical framework to explain the magnetic reconnection in oppositely directed magnetic field lines in plasma and formation of the current sheet. However, the reconnection rate given by this model is much slower than those observed in the solar corona. The Sweet-Parker model was modified by reducing the size of resistive layer in the reconnection region thereby speeding up the reconnection rate \citep{Petschek1964}. 
A simplified 2D model, CSHKP, was put forward to explain the flares and related observed phenomena such as flare ribbons and post-flare loops \citep{Carmichael1964, Sturrock1966Nature, Hirayama1974SoPh, Kopp1976SoPh}. This was extended to a plasmoid induced reconnection model by \cite{Shibata1996AdSpR, Shibata1997ESASP}. 
It was also suggested that the magnetic reconnection plays a major role in the CME acceleration. It was proposed that in an isothermal atmosphere, the average Alfv\'en Mach number (M$_A$) for the inflow in to the reconnection region, as low as 0.005, is sufficient to cause an eruption \citep{Lin2000JGR}. This limit was increased to 0.013 taking into account a more realistic plasma atmosphere \citep{Sittler1999ApJ, Lin2002ChJAA}.
\cite{Shibata2001EP&S} showed that the ejection of plasmoids from current sheet leads to a highly time dependent faster reconnection at different spatial and temporal scales. \cite{Vrsnak2003SoPh} estimated that during the onset of the fast reconnection process, the ratio of width to length of the current sheet lies in a range of 1/18 to 1/9 leading to tearing mode instability and formation of plasmoids. 
The role of plasmoid instability in the fast reconnection rate was reported in \cite{Bhattacharjee2009, Huang2010}. 
Theoretically, it is proposed that in the current sheets with high Lundquist numbers, the distribution function $f(\psi$) of the magnetic flux $\psi$ associated with plasmoids follows a power law, $f(\psi$) $\sim$ $\psi^{-1}$ \citep{Huang2012PhRvL}. However, due to lack of magnetic field measurements in corona, this could not be verified observationally.
{ In an ultra-high resolution 2.5D simulation \citet{Guidoni_2016ApJ} studied the evolution of magnetic islands formed in the current sheet and found that the contraction of these islands may provide enough energy to accelerate particles.}
The plasmoids formed in a gravitationally stratified current sheet during the reconnection process move upwards and downwards.  Moreover, small plasmoids merge to form larger plasmoids that show oscillatory behaviour in their size as they propagate. Furthermore, downward propagating plasmoids cause the underlying post flare loops to oscillate upon interaction \citep{Tom2017ApJ}.

The theoretical models have always been improved to support the observations. During the 2002 July 23 solar flare, it was found that the downflow motions are correlated with the release of the magnetic energy \citep{Asai2004ApJ}. The observations of flows in post-CME current sheets are limited by the resolution of the current instruments. \cite{2010cartwheel} (and references there in) reported the observations of Supra-arcade downflows (SADs), observed as dark density depleted structures moving sun-ward, associated with flare leading to CME eruption using multi-wavelength data. 
On the other hand, plasmoids are bright density enhanced features observed as blobs in the current sheet identified in white-light images of Large Angle and Spectrometric Coronagraph  (LASCO) \citep{Brueckner95} and have been reported by \cite{Ko_cs_2003, riley_plasmoids2007ApJ,Schanche2016ApJ, lasco_cs_stats2016SoPh, Chae2017ApJ}. 
The observations of slowly drifting structures in radio wavelengths also provided signatures of downward moving plasmoids along the current sheet \citep{Kliem2000A&A, Ning2007SoPh}. These are associated with the secondary tearing happening in the current sheet and the coalescence of smaller plasmoids to form the larger ones. Evidence of bidirectional plasmoids during the reconnection using multi-wavelength observations were reported in \citet{kumar2013, Takasao2012ApJ}.
In recent studies of the observed current sheet of September 10, 2017, downflows have been observed and associated with the reconnection \citep{Longcope2018ApJ, qpo2019ApJ, Lee2020ApJ, yu2020magnetic}.

The scales of the observed moving structures associated with reconnection have been analysed in recent studies in order to understand the spatial and temporal scales present. A statistical study of SADs observed during flares revealed that the size of SADs follow a log-normal distribution where as the flux follow both, a log-normal and an exponential distribution \citep{McKenzie2011ApJ}.
In another statistical study using the LASCO observations, plasmoid sizes were found to have two distributions. The number of blobs first increase and later decrease with increase in the widths \citep{guo_plasmoids2013ApJ}. It was also compared with a resistive MHD simulation and the possibility of a power-law distribution was suggested which differ from the log-normal distribution as proposed by \citet{McKenzie2011ApJ} for donwflowing structures.

Apart from the size of plasmoids, their kinematical properties have been reported in various studies.
When the blobs in the LASCO images were observed with an average speed ranging from 300-650 km s$^{-1}$, it was suggested that the plasmoids move with local Alfv\'en speed in the current sheet after the reconnection \citep{Ko_cs_2003}. For the 2003 November, 9, a current sheet was identified in the wake of the CME along which 5 blobs were observed in the LASCO FOV. These blobs were recorded with an average speed range of 460 - 1075 km s$^{-1}$ \citep{Lin_2005}.
Employing a 2-D reistive-MHD simulation for magnetic reconnection, it was reported that the upward moving plasmoids gain the highest speed of the order of ambient Alfv\'en speed (V$_A$) where as the downward moving plasmoids attain only fraction of this speed \citep{Barta2008A&A, Forbes2018ApJ}. 
In a study combining EUV and radio observations, bidirectional plasmoids formed during the reconnection were observed to be moving with speeds of $\sim$152 - 362 km s$^{-1}$ and $\sim$83 - 254 km s$^{-1}$, respectively \citep{kumar2013}.
An average speed of 307 km s$^{-1}$ with a spread from 93 to 723 km s$^{-1}$ was derived using LASCO/C2 images by identifying and tracking 9 upward moving blobs associated with different CMEs \citep{Schanche2016ApJ}.
In a recent statistical study, the blobs were identified in LASCO coronagraph images over minima (1996-98) and maxima (2001) of solar cycle 23. Their speed range was found to be 245-462 km s$^{-1}$ associated with $\sim$130 post-CME current sheet(/rays) \citep{Webb2016SoPh}. It was reported in \citet{Forbes2018ApJ} that due to limited observations, the upward moving blobs do not show variation in speed along the current sheet whereas deceleration was observed for downward directed blobs. This was unlike the kinematic behaviour of plasmoids obtained from their simulation.
In a candidate current sheet associated with a CME observed by COR-2 on board {\it Solar TErrestrial RElations Observatory} (STEREO) \citep{Howard02}, the plasma blobs were tracked using an automated method and were found to have average speed of 303 km s$^{-1}$ with the starting height of blob formation as 0.5 R$_\odot$ above the solar surface \citep{Chae2017ApJ}. 

In spite of several statistical and theoretical studies we do not have a clear understanding of the plasmoids evolution during the reconnection process. 
In this study we investigate the properties of plasmoids observed in post-CME current sheet in multi-wavelength data on September 10, 2017. We analysed the size distribution of the observed plasmoids in different pass-bands extending the lower limit which shows a power-law distribution rather than log-normal \citep{guo_plasmoids2013ApJ}. We also studied their speed distribution over different heights in the solar corona and try to correlate them with their size during propagation. We identified the primary disconnection point in the current sheed using the information of the direction of motion of the plasmoids.
The instruments used for the observed data are presented in Section \ref{sec:Observation}. We describe the procedure followed to derive the Alfv\'en Mach number for the current sheet, identification of plasmoids and the results of the statistical properties of the observed plasmoids along with determining the disconnection location in the current sheet in Section \ref{sec:analysis}. Section \ref{sec:summary} summarises our interpretation of the results followed by discussion.

    

\section{Observations}
\label{sec:Observation}
A X8.3 flare occured on September 10, 2017 at NOAA Active Region (AR) 12673 near the west limb of the Sun and is associated with a fast CME of speed 3200 km s$^{-1}$ \citep{NatG2018cs, astridcs2018ApJ...868..107V}. This event also lead to the formation of a current sheet at $\sim$16:10UT  above the location of a flare for a very long time and has been reported in recent studies \citep{micro_hardx_2018ApJ, broadening_cs2018ApJ, cs_turbulence2018, French2019ApJ, fluxrope2018ApJ, spectral_cs2018ApJ, qpo2019ApJ, yu2020magnetic, Lee2020ApJ}.

We used Extreme Ultraviolet (EUV) images of {\it Atmospheric Imaging Assembly} (AIA) onboard {\it Solar Dynamics Observatory} (SDO) \citep[AIA;][]{AIA} taken at 131 \AA\ pass-band for high resolution observation near the solar limb starting from 15:30UT to 19:00UT. These images with a pixel resolution of 0.6 arcsec were reduced to level 1.5 using the standard IDL routine, {\it aia\_prep.pro}, to correct for the rotation, alignment, and adjust the plate scale. We used images having exposure time of $\sim$2.9 s with cadence of 24 s. 
The current sheet is also observed in white-light coronagraph images of the K-Cor telescope at Mauna Loa Solar Observatory (MLSO) \citep{deWijn} and {\it Large Angle Spectroscopic Coronagraph} (LASCO) onboard {\it Solar and Heliospheric Observatory} (SOHO) \citep{Brueckner95}. We used the available K-Cor level 1 images processed with a Normalised Radial Gradient Filter (NRGF) \citep{NRGF2006SoPh} having field of view (FOV) from 1.05 R$_\odot$ to 3 R$_\odot$ at a cadence of 2 minutes and resolution of 5.6 arcsec pixel$^{-1}$ from 17:00UT to 19:00UT.
 
\begin{figure}[htbp]
   \centering
   \includegraphics[angle=0,width=9cm]{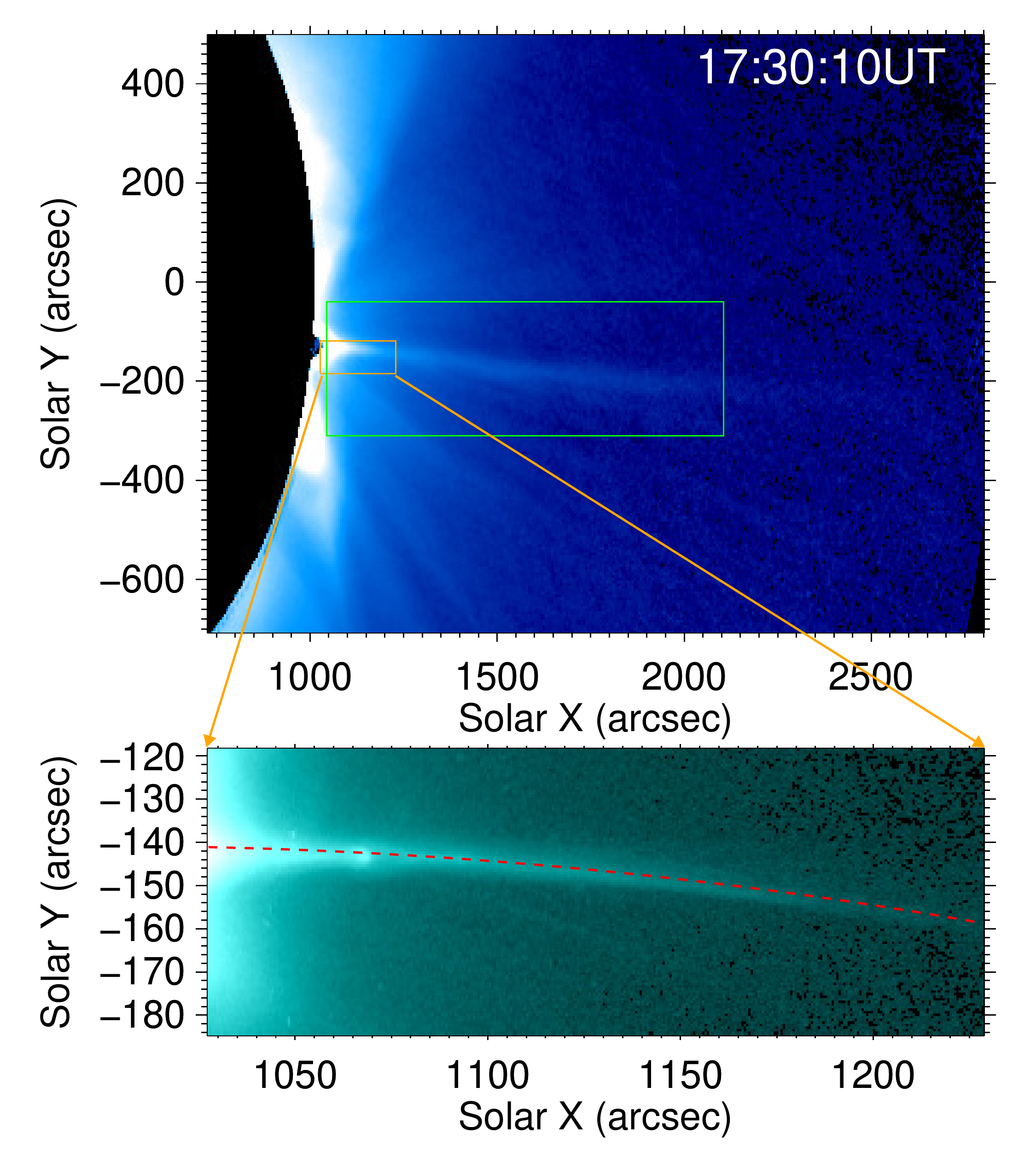}
      \caption{Context image for the current sheet in the K-Cor data. Green rectangular box in the {\it Top} panel corresponds to ROI-A chosen for the analysis in the K-Cor data. Orange box represents the ROI-B chosen in the AIA 131 \AA\ passband data, which is shown in the {\it Bottom} panel. The red dashed line along the identified current sheet is used for generating height-time plot.
       (An animation is available for this figure)}
         \label{fig: context_kc}
   \end{figure}

The observed current sheet is show in Figure \ref{fig: context_kc} where the top panel shows a K-Cor observation while AIA 131 \AA\ observation for a smaller FOV is shown in the bottom panel. The Region of Interest (ROI) was chosen along the current sheet which is highlighted by a green box in the K-Cor image (ROI-A). A smaller region in the same image shown in orange box corresponds to the ROI in AIA 131 \AA\ images (ROI-B). 
The red dashed line along the current sheet in the bottom pannel of Figure \ref{fig: context_kc} is used to generate height-time plot.
A careful observation of ROI-B in successive AIA images revealed some of the plamoids moving upward while others moving downward towards the Sun along the current sheet. The signature of plasmoids is also noted in the K-Cor images. An animation is available for Figure \ref{fig: context_kc} showing the identified plasmoids in the K-Cor FOV. Moreover, we used LASCO/C2 images with cadence of 12 minutes and pixel resolution of 11.4 arcsec pixel$^{-1}$ to track the outward moving plasmoids up to 6 R$_\odot$. The images from LASCO/C3 coronagraph, having FOV from 4 to 30 R$_\odot$, and spatial resolution of 56 arcsec pixel$^{-1}$ resolution were used to observe the current sheet further into the corona.

\section{Analysis and Results}
\label{sec:analysis}

Images of AIA 131 \AA\ were radially filtered using {\it aia\_rfilter.pro} available in Solarsoft package of IDL, enhance the off-limb features. This resulted in visualisation of the current sheet up to $\sim$1.3 R$_\odot$. It also enhanced the brightness of anti-sunward moving plasmoids/plasma blobs up to larger distances. Henceforth, anti-sunward and Sun-ward moving plasmoids are referred as upward and downward moving plasmoids respectively.
The features in the EUV difference images were further enhanced with intensity scaling for better visualisation. 
LASCO/C2 and C3 images were also processed with a radial filter to identify the features throughout their FOVs.

\subsection{Estimation of the Alfv\'en Mach Number}

  \begin{figure*}[h]
   \centering
    \vspace{-2.5cm}
   \includegraphics[angle=0,width=17cm]{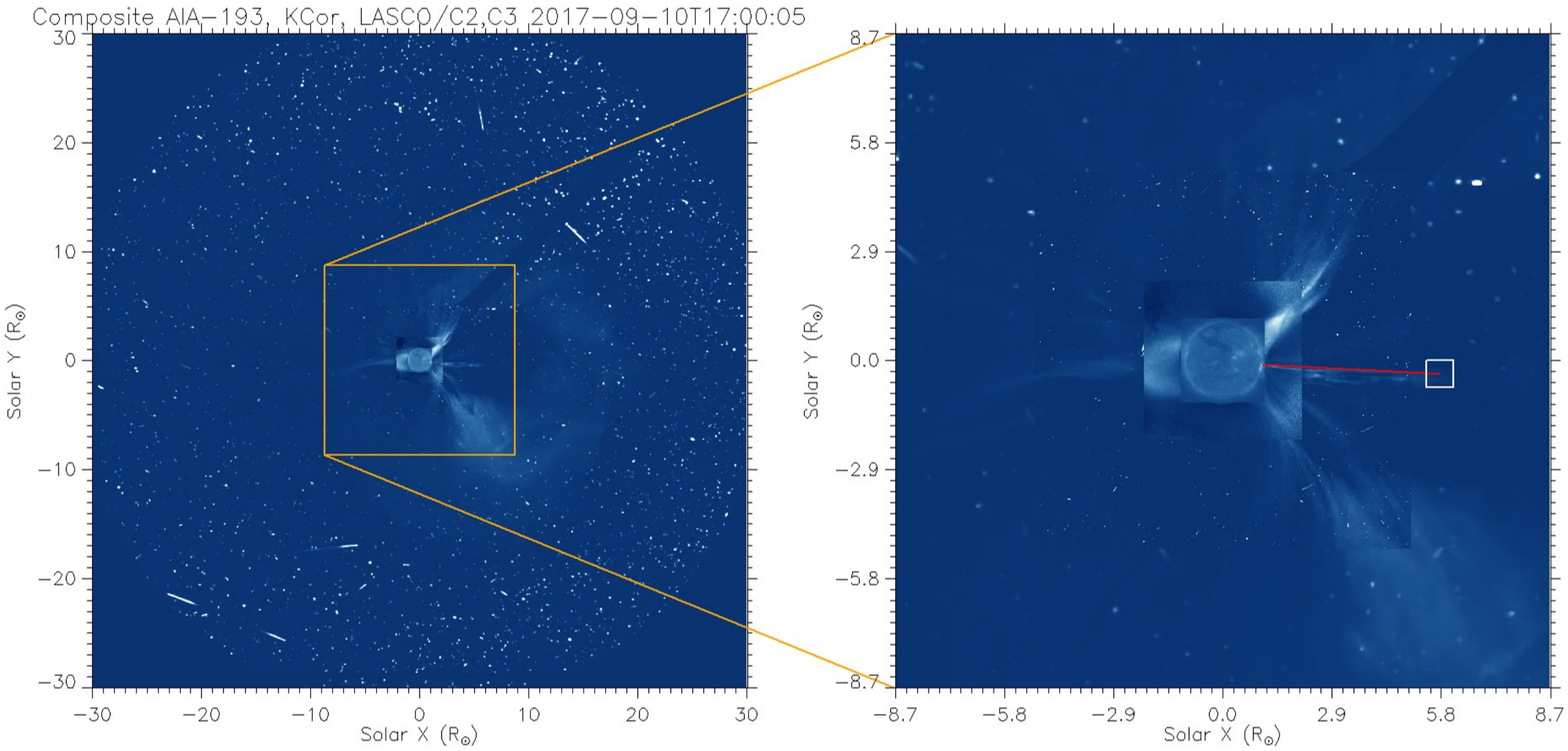}
   \vspace{-1.5cm}
      \caption{Composite image of AIA 193 \AA\ , K-Cor, LASCO-C2 and C3 for September 10, 2017 at 17:05 UT. A close view of the inner FOV is shown in subsequent image with the length of current sheet shown by the red line. White rectangular box represents the approximate error in estimating the length of the current sheet.}
         \label{fig: mach}
   \end{figure*}
   
It can be seen in Figure \ref{fig: context_kc} that the identified current sheet is visible in pass-bands extending from EUV to white-light.
We estimated the length of the current sheet in inner corona using AIA 131 \AA\ images and further in white-light coronagraph images extending to outer corona using K-Cor and LASCO coronagraphs following the procedure as reported in \citet{Lin2000JGR, 2010cartwheel}. Assuming that lower end of the current sheet (p) remains fixed over the post-flare loops ($\sim$1.03 R$_\odot$), the length of the current sheet is estimated to from p to the height observed (q) in AIA, K-Cor and LASCO coronagraphs composite image as shown in Figure \ref{fig: mach} where the red line shows the length of the current sheet. The box at the its end represents the uncertainty in measurement of its length. Multiple measurements are taken to increase the accuracy of the length of the current sheet.
The average width of the current sheet measured in EUV pass-band using AIA 131 \AA\ images is estimated to be $\sim$9 Mm between the height of 1.03 to 1.1 R$_\odot$.
The average width of the current sheet in white-light images of LASCO/C2 is measured to be $\sim$ 90 Mm below 2.5 R$_\odot$ whereas in K-Cor images is is estimated to be  $\sim$ 20 Mm averaged over height from 130 to 390 Mm. 
The width measurements are close to the values reported by \citet{cs_turbulence2018}  as $\sim$10 Mm in EUV images while $\sim$ 25 Mm in white-light (K-Cor) for the same current sheet. The measured width of the current sheet is wider than the theoretical predictions due to the fact that we are making measurements based on the density of plasma surrounding the structure. Furthermore, the thermal conduction in the reconnection region heat the surrounding plasma and are responsible for slow mode shocks and flows along the current sheet which leads to identification as wider current sheet \citep{Seaton2009ApJ}.

\begin{figure}[!h]
   \centering
   \includegraphics[angle=0,width=9cm]{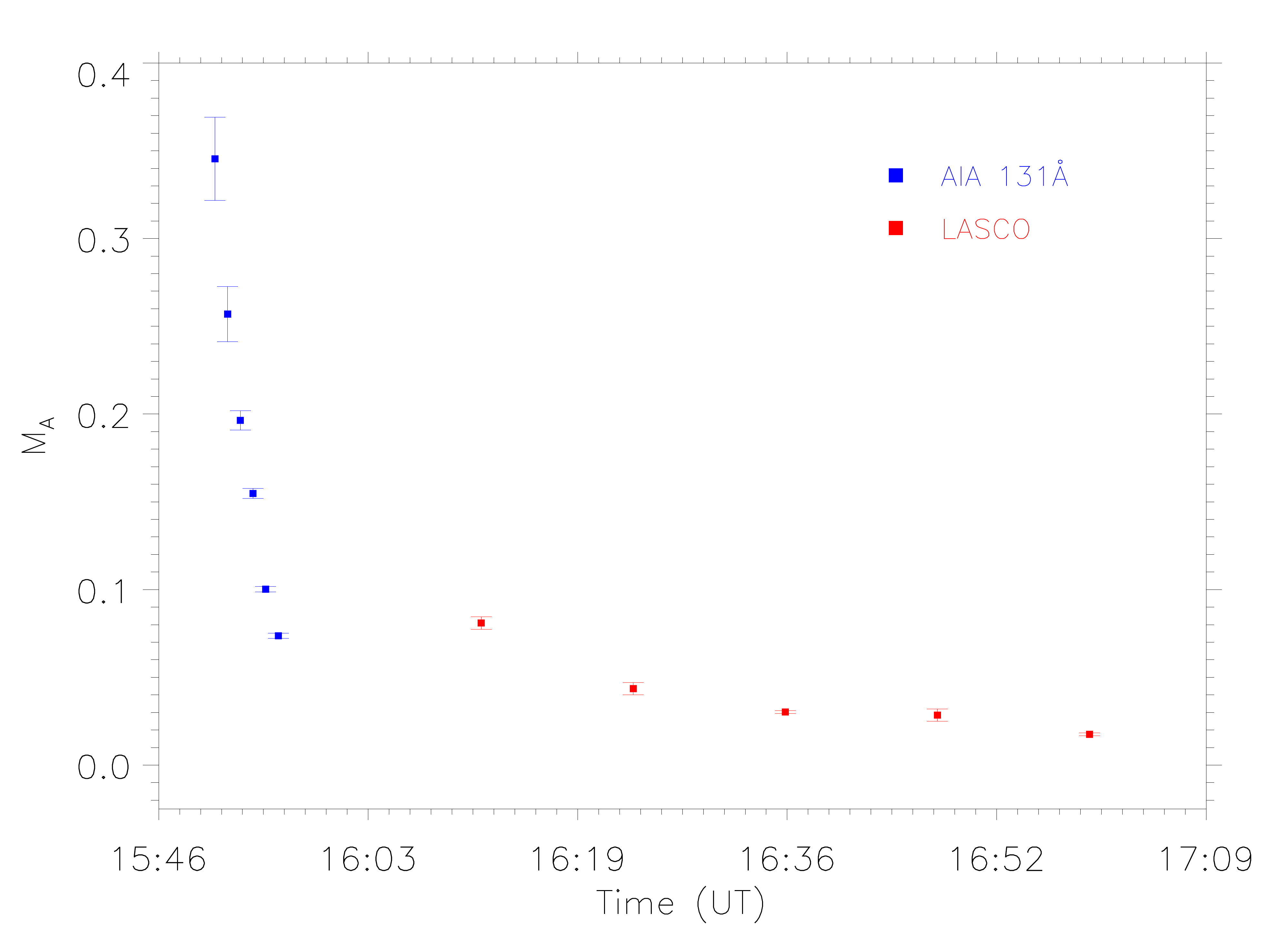}
      \caption{Evolution of estimated Alfv\'en mach number (M$_A$) with time using the relation width/length of current sheet. The blue points are the M$_A$ estimated using AIA 131 \AA\ images for current sheet length measurement where as the red ones belong to LASCO images.
    }
         \label{fig: machplot}
   \end{figure}

The Alfv\'en mach number (M$_A$) is estimated as the ratio of width and length of the current sheet where the length is determined by q-p \citep{Lin2000JGR, 2010cartwheel}. The value of M$_A$ is plotted with time in the observed current sheet as shown in Figure \ref{fig: machplot}. The points in blue are the measurement made using AIA 131 \AA\ images in inner corona where as the red points correspond to estimation with LASCO images. The jump in the estimated values after 16:00UT may be due to the fact that two different passbands have been used here for the estimation of dimensions of the current sheet.
The values obtained range from 0.018 to 0.35 which satisfy the lower limit of M$_A$ proposed by \cite{Lin2000JGR, Lin2002ChJAA} for an eruption to occur. 
M$_A$ near the reconnection site also represents the rate of reconnection. The reconnection rate of 0.003-0.2 has been reported in \cite{cs_turbulence2018} estimated by taking the ratio of inflow velocity to the outflow velocity at the reconnection site using AIA 193 \AA\ images for this event. It should be noted that as K-Cor images are available from $\sim$ 17:00 UT when the current sheet had already reached LASCO/C2 FOV. 
The M$_A$ for this event is higher that the values obtained for the current sheet reported by \cite{2010cartwheel} as 0.002-0.006, comparable to the earlier estimated values reported in \cite{Ko_cs_2003, Lin_2005, Bemporad2006ApJ, Ciaravella2008ApJ, Takasao2012ApJ}, and lower than the range of 0.15-0.27 estimated for a novel forced reconnection observed on 2012 May 3 \citep{Srivastava2019ApJ}.

\subsection{Identification of Plasmoids in Current Sheet}

 \begin{figure*}[ht]
   \centering
    \includegraphics[angle=0,width=18cm]{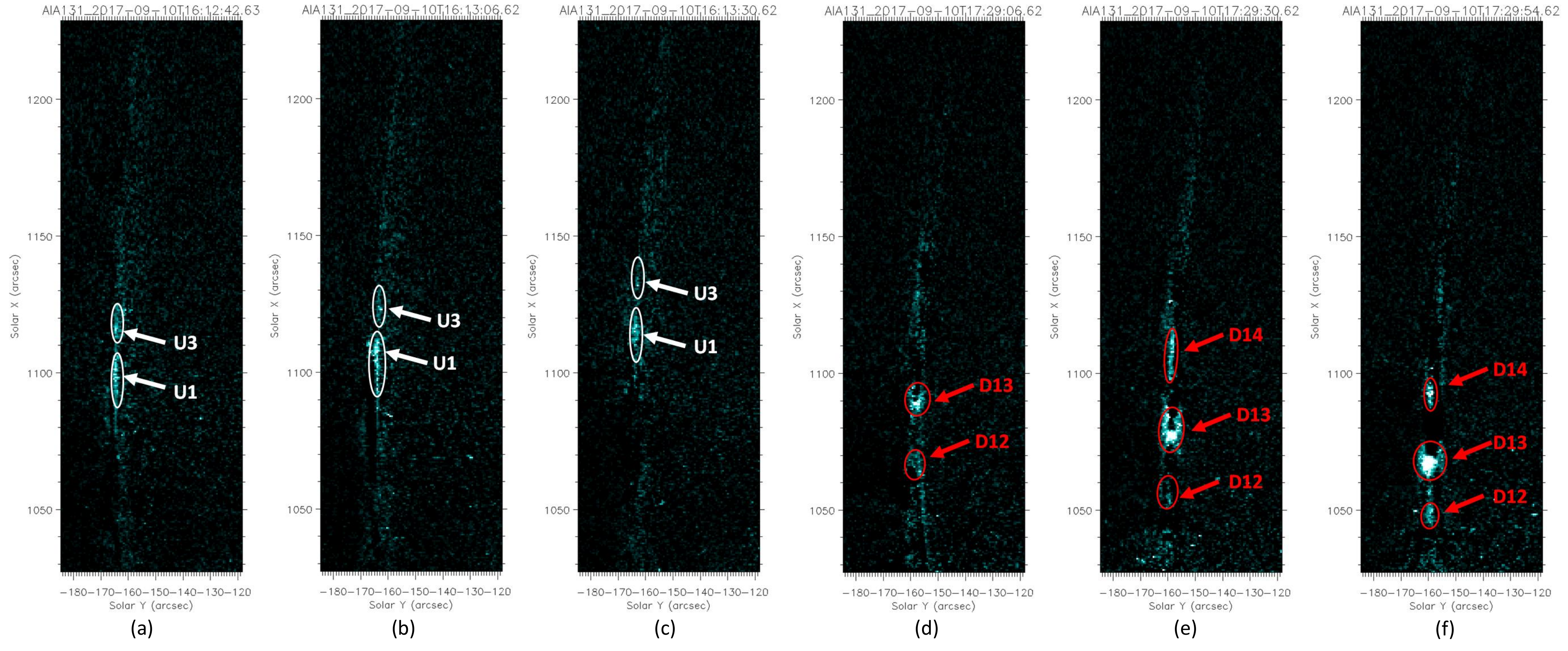}
    \includegraphics[angle=0,width=18cm]{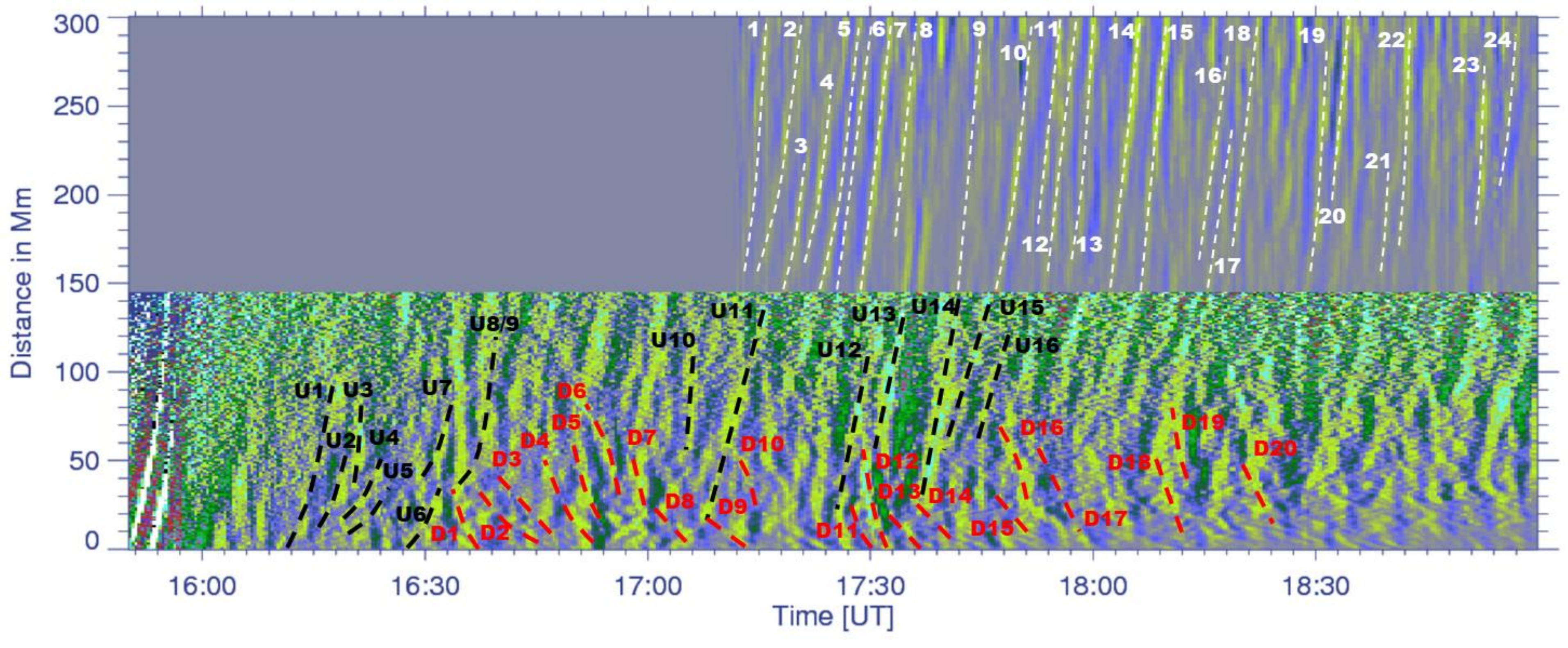}
       \caption{{\it Top} Plasmoids identified in radial filtered difference images of AIA 131 \AA\ along the current sheet. The outward and downward moving plasmoids are shown from (a) to (c) and (d) to (f) panels encircled in white and red colors respectively. The {\it Bottom} pannel shows the height-time plot corresponding to Figure \ref{fig: context_kc} having AIA FOV upto $\sim$150 Mm and K-Cor beyond it. The ridges corresponding to the identified blobs are marked by dashed lines in black and red colors respectively in AIA FOV and in white within K-Cor FOV. The serial number of the plasmoid in the top panel have their corresponding ridges in the bottom height-time plot.
       (An animation is available for this figure)}
      \label{fig: sample_updown}
   \end{figure*}

It should be noted that the estimated M$_A$ is satisfies the requirement proposed by \cite{Vrsnak2003SoPh} for the formation of plasmoids during the reconnection process. We found that there were plasmoids along the current sheet in both EUV and white-light observations.
Plasmoids were identified by visual inspection as discrete intensity enhancements moving along the current sheet in AIA images taken at 131 \AA\ and white-light images of K-Cor and LASCO/C2. { In general, plasmoids are characterised by O point and hence termed as the magnetic islands. Thus, using only intensity images, it is difficult to distinguish plasmoids from flows. Therefore, we have to rely on the shape and thus we classified a near-elliptical and elongated confined intensity structures as the plasmoids in contrast to continuous plasma motion in case of flows.} We inferred the direction of motion of plasmoids after tracking them in consecutive images.
It is important to note that plasmoids are identified as brightness enhancement due to increased density. These are different from SADs reported in \citet{2010cartwheel, McKenzie2011ApJ} which are identified as depleted density and hence observed as dark features in successive frames.
Figure \ref{fig: sample_updown} shows an example of such plasmoids. We have manually put contours around these structures by observing the significant change in the intensity around the plasmoids taken as the edges. { The upward moving plasmoids are outlined in white color contours while downward moving are outlined in red. The same have also been marked in Figure~\ref{fig: sample_updown} (and its associated animation available online) using AIA 131 \AA\ images. The animation consists of intensity images on the left side and radial filtered difference images with scaled intensities on the right. Upon careful inspection the blobs could also be identified in the original intensity images at the locations marked in difference images in Figure~\ref{fig: sample_updown}.} The temporal location of these plasmoids were also verified with the help of height-time plot shown in the middle panel of the same figure. { It should be noted that the bidirectional blobs are produced at few instances which could be easily seen as discrete propagating structures in either direction. However, these signatures are not always visible which may be due to the limitations of the visual identification, line of sight effetcs or geometry of the current sheet. The nature of plasmoids as seen from height-time plot is also similar to those generated in the MHD simulation by \citet{Guidoni_2016ApJ}.}

\begin{figure*}[!h]
   \centering
   \includegraphics[angle=0,width=18cm]{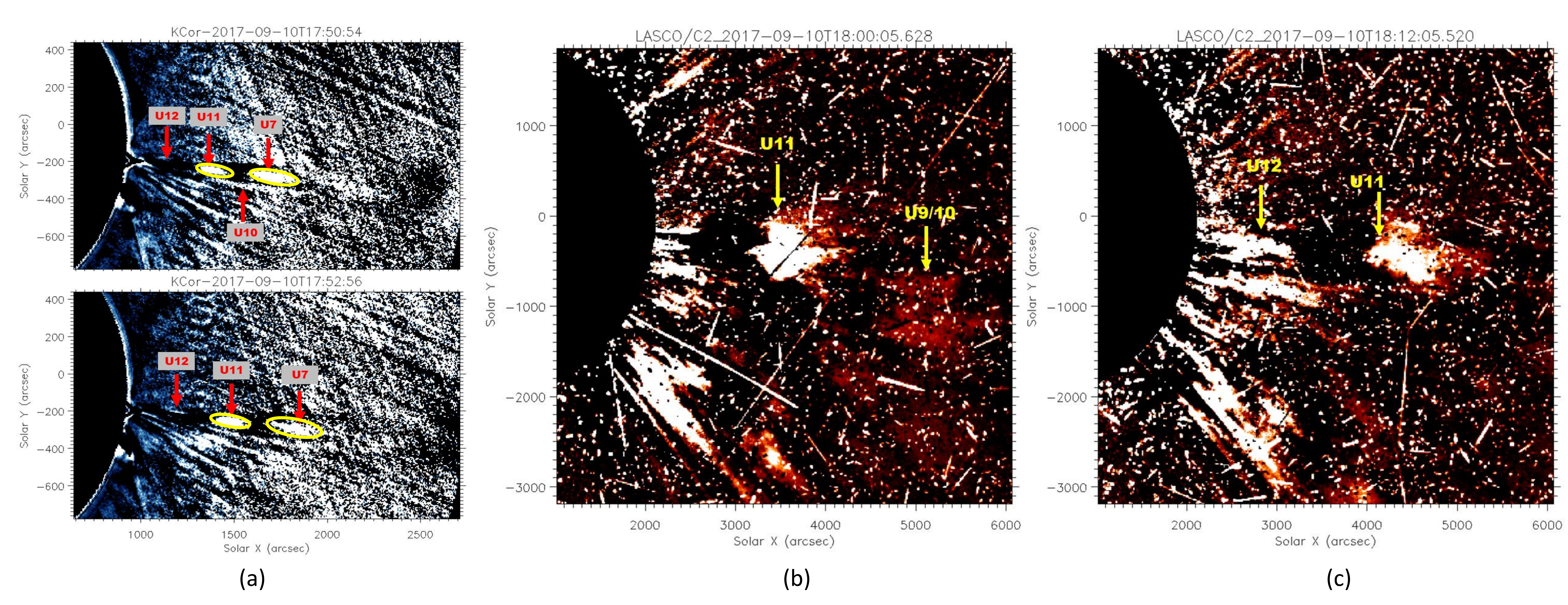}
      \caption{Identification of plasmoids in successive frames of white-light coronagraph images, {\it (a)} In K-Cor difference image plasmoids are marked with yellow contours, {\it (b) and (c)} In LASCO/C2 difference image the location of identified blobs are pointed with yellow arrows.
      (An animation is available for this figure)}
         \label{fig:whiteblobs}
   \end{figure*}

Similar to the identification of plasmoids in EUV, we identified them as discrete elongated structures in the running difference images with intensity scaling of K-Cor and LASCO/C2 as shown in Figure \ref{fig:whiteblobs}. For this pass-band of observation, only upward moving plasmoids were identified. It may be because these plasmoids are smaller in size and form at lower heights that are unresolved by these instruments. An animation available in online version for Figure \ref{fig:whiteblobs} shows the identified plasmoids in the LASCO/C2 FOV.
It can be seen in the difference images, the plasmoids are distinctly visible at the location where current sheet is seen. We would like to mention here that K-Cor being a ground based coronagraph suffers from atmospheric contribution, hence, not all the plasmoids could be tracked in successive frames as those for AIA and LASCO images.
We recorded plasmoids after their first appearance in AIA 131 \AA, and LASCO/C2 images. In K-Cor images first appearance of most of the plasmoids were identified between 1.2 to 1.5 R$_\odot$ when they can be identified as moving structures. This is due to lower contrast of the plasmoids with respect to the background and bigger size probably due to falling density and magnetic field of the ambient medium with radial distance. This points towards the fact that the disconnection point must be lower than 1.2 R$_\odot$. This also suggest that the downward moving plasmoids could not be identified in KCor images as they formed at heights lower than 1.2 R$_\odot$ which has been mostly difficult to observe in these images due to limited resolution and saturation close to the flaring region.

From the inspection of AIA 131 \AA\ images, we found that there were 20 downward moving plasmoids whereas 16 upward moving plasmoids. On the other hand 24 upward moving plasmoids were identified from K-Cor images in the inner corona for the same time period whereas using LASCO/C2 images we found 17 plasmoids moving out in the outer corona for an extended period up to 20:00UT. The discrepancy in the lower number of upward moving bobs in AIA FOV may be due to the limitation in the manual identification as most of these were observed towards the outer region where the signal to noise becomes relatively low.
 We also recorded the position of these plasmoids in successive images of AIA, K-Cor, and LASCO/C2 images to estimate their kinematic properties. 

  
    
   
\subsection{Distribution of Plasmoids}

\subsubsection{Size Distribution}
\begin{figure}[!h]
   \centering
   \includegraphics[angle=0,width=9.5cm]{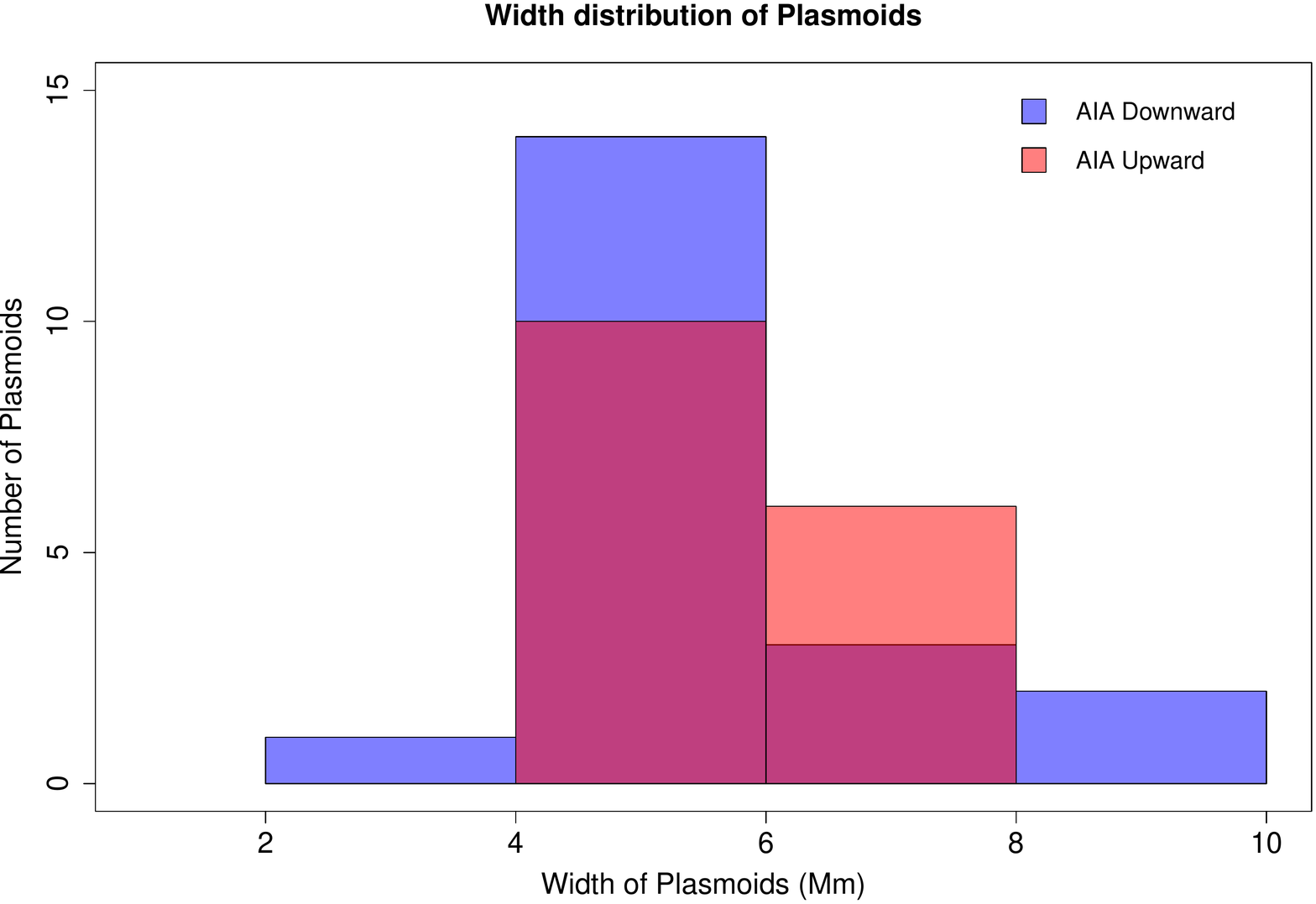}
   \includegraphics[angle=0,width=9.5cm]{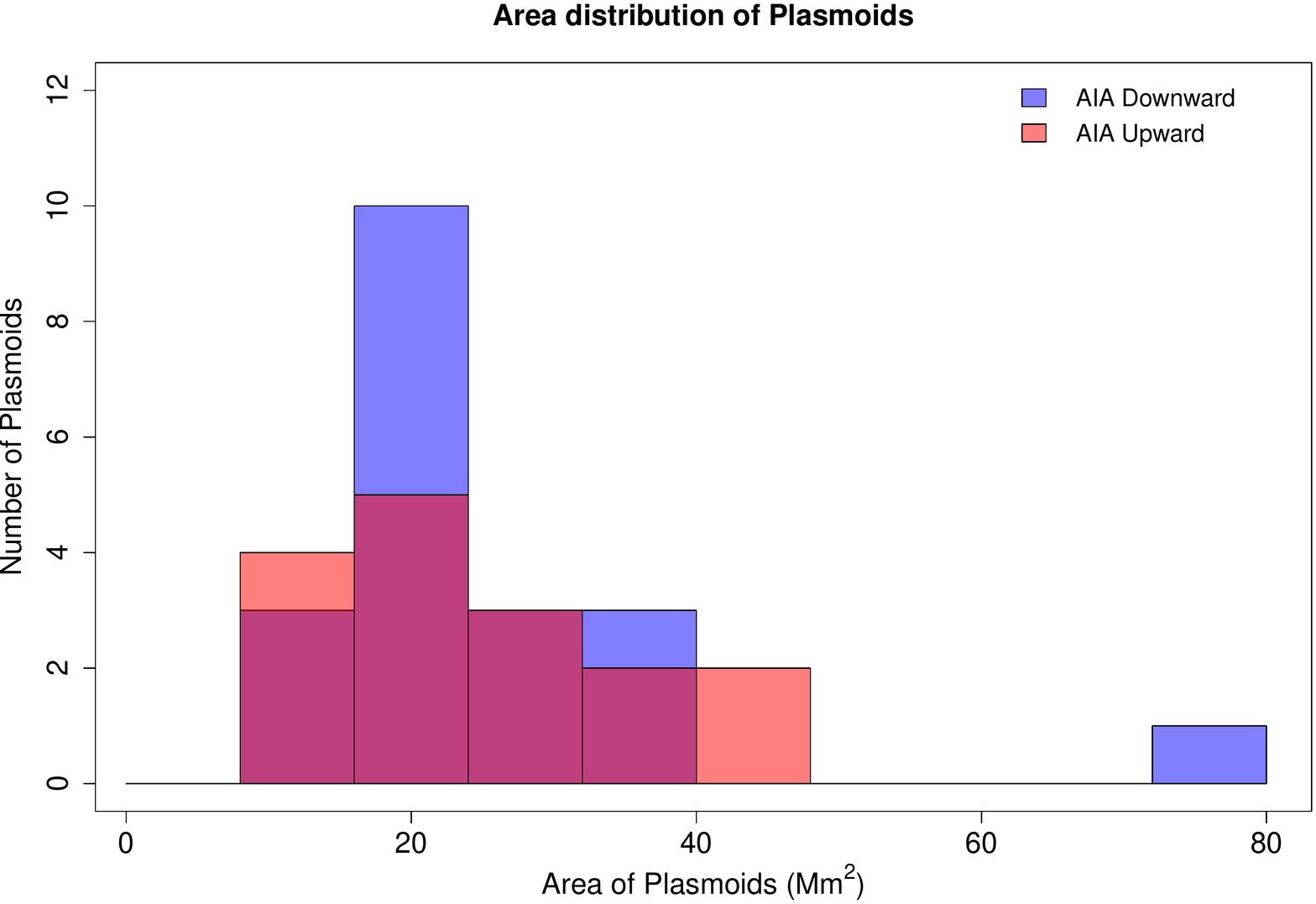}
      \caption{Histogram of the size distribution of the blobs identified in AIA 131 \AA\ passband images, {\it Top} Width distribution of blobs, {\it Bottom} Area distribution of blobs.}
         \label{fig:size_hist_131}
   \end{figure}

We noted the length of major and minor axes of the plasmoids when they first appear in the FOV assuming that they posses near-elliptical shape. The width of the plasmoids were then estimated as an average of these two dimensions. 
Figure \ref{fig:size_hist_131} shows the size distribution of plasmoids identified in AIA 131 \AA\ images where the top panel shows the average width and the bottom pannel the calculated area of plasmoids considering ellipse like shape. The two colors show the  distribution of upward and downward moving plasmoids respectively. We found that the upward moving plasmoids show widths ranging from 4 to 8 Mm with maximum number of plasmoids (10 out of 16) having widths in the range 4 to 6 Mm. The average width of these plasmoids is 5.65 Mm with median value of 5.28 Mm. On the other hand, the downward moving plasmoids have distribution from 2 to 10 Mm and maximum plasmoids (14 out of 20) have widths in the range of 4 to 6 Mm. These plasmoids have average width of 5.92 Mm and median width of 5.65 Mm. 
The area distribution of these plasmoids as shown in lower pannel of Figure \ref{fig:size_hist_131} shows a wide range of measured area. The area of upward moving plasmoids ranges from 10 to 50 Mm$^{2}$ with average and median area of 23.93 and 21.29 Mm$^{2}$ respectively. The downward moving plasmoids showed a larger range of area from 10 to 80 Mm$^{2}$ having average area of 25.89  Mm$^{2}$ and median of 22.75  Mm$^{2}$. This shows that both the upward and downward moving plasmoids show a similar size distribution when identified first in AIA FOV. The small differences may be due to the limitation and subjectivity of the visual inspection.

\begin{figure}[!h]
  \centering
  \includegraphics[angle=0,width=9cm]{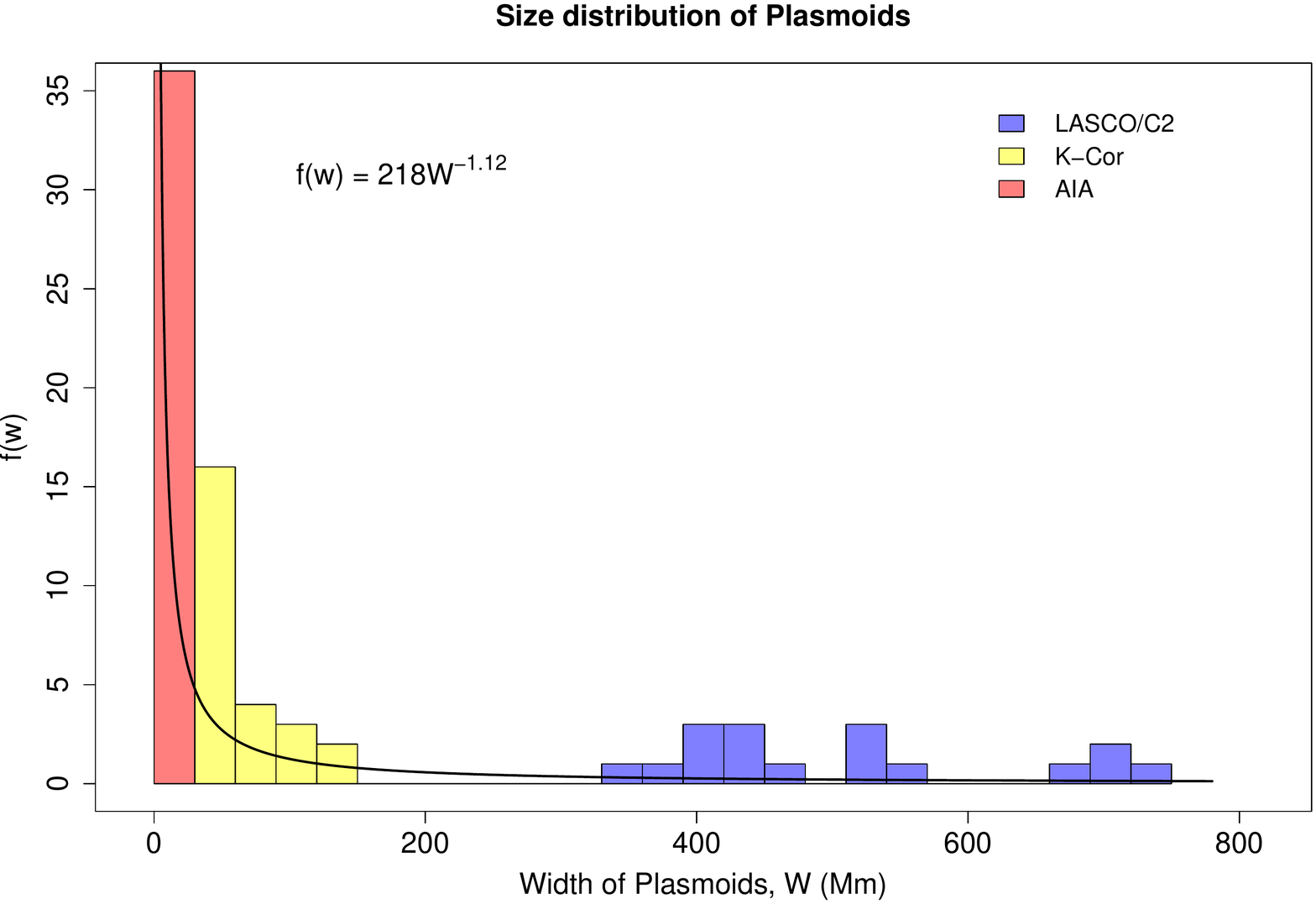}

      \caption{Width distribution of plasmoids identified in AIA, K-Cor and LASCO C2 images. The horizontal axis is the measured width when they are firest identified in the FOV, while the vertical axis is the number of plasmoids corresponding to each bin. The black solid curve is the power law fitting to represent the distribution.}
         \label{fig:size_all}
  \end{figure}

The width and area distribution of the plasmoids identified in AIA FOV appears to show a power-law with cut-off as can be seen from Figure \ref{fig:size_hist_131}. This cut-off may be due to the limited resolution of the AIA instrument to observe the smallest plasmoid followed by manual subjectivity.
Therefore, we analysed the overall size distribution of plasmoids from inner to outer corona combining the information about plasmoids width observed in AIA 131 \AA\ , K-Cor, and LASCO/C2 as shown in Figure \ref{fig:size_all}. The bin corresponding to AIA includes all the upward and downward moving plasmoids having mean width of 5.8 Mm. The plasmoids observed in K-Cor images have widths from 30 to 150 Mm with average width of 64 Mm and have median width of 54 Mm. Further, LASCO/C2 observations shows larger range of observed widths. The plasmoids had width ranging from 320 to 750 Mm with average and median width of 510 and 456 Mm respectively. It can be seen from Figure \ref{fig:size_all} that the size of plasmoids increases from inner to outer corona. It was noticed that average width of plasmoids increases by $\sim$11 times when they evolve from AIA to K-Cor FOV which further increase by $\sim$8 times on reaching outer corona in LASCO/C2 FOV. It turns out that the plasmoids show a heavy-tail distribution of their sizes when the three intrument observations are combined.
We fitted the size distribution by a power law with maximum likelihood \citep{Clauset2007} giving the relation,
\begin{equation}
    \label{eq:pw_size}
    f(W) = 218W^{-1.12},
\end{equation}
where f(W) is the number of plasmoids with width W. 
\citet{guo_plasmoids2013ApJ} using only LASCO/C2 images observed a dip in the size distribution of plasmoids below 50 Mm extending the possibility of missing the smaller ones by visual inspection. However, combining the observations from inner to outer corona, we find that the size distribution of plasmoids formed during the reconnection process follow a single power-law distribution in small as well as large scale regimes which was also proposed by their simulation. This is the first observational evidence in the support of this model representing the size distribution of plasmoids by a single power-law.

   
   
\subsubsection{Speed Distribution}

\begin{figure}[!h]
\centering
\includegraphics[angle=0,width=9cm]{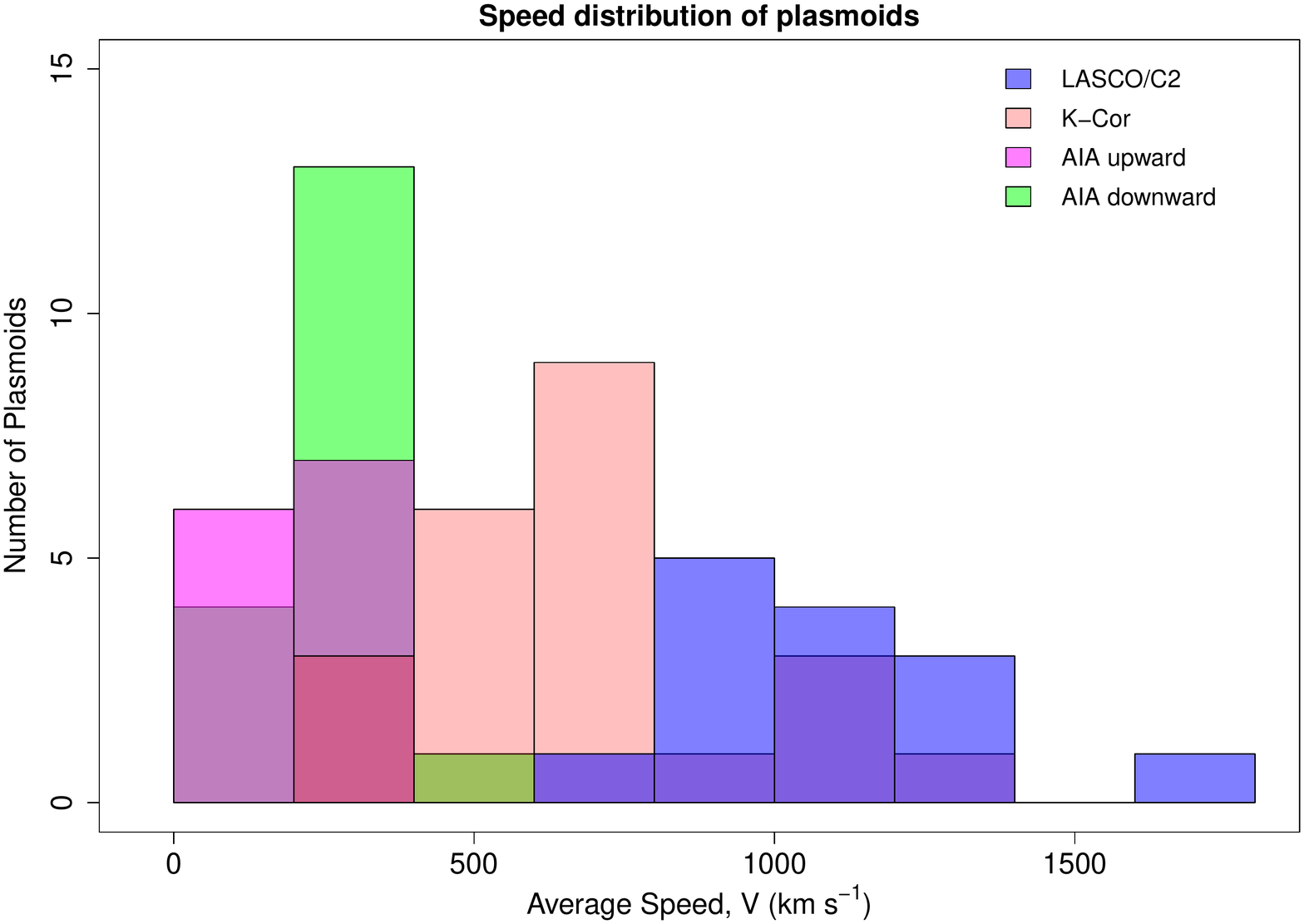}

  \caption{Speed distribution of all plasmoids observed from inner to outer corona. Different colors represent observations from different instruments used for this study.}
     \label{fig:speed_dist_all}
\end{figure}

We tracked the blobs in subsequent images and recorded their positions. Most plasmoids could be identified in at least two successive images. We were able to derive the average speed of the plasmoids observed in EUV as well as white-light images. Figure \ref{fig:speed_dist_all} shows the distribution of average speed for all the plasmoids which could be tracked. We found that the downward moving plasmoids identified in AIA 131 \AA\ images had speed ranging from 70 to 425 km s$^{-1}$ with average speed of 272 km s$^{-1}$ where as the upward moving plasmoids in same FOV had speed range of 30 to 400 km s$^{-1}$ with mean at 191 km s$^{-1}$. It can also be seen from Figure \ref{fig:speed_dist_all} that for both upward as well as downward moving plasmoids, most of the plasmoids fall in the speed range of 200 to 400 km s$^{-1}$. 
The upward moving plasmoids were difficult to track in K-Cor images due to poor SNR. The NRGF processed K-Cor images provided a better performance for tracking the plasmoids at heights from 1.2 to 2 R$_\odot$ with a lower cadence of 2 minutes compared to 15 seconds of the prior. The plasmoids moved upwards with speed ranging from 320 to 1360 km s$^{-1}$ having an average speed of 671 km s$^{-1}$. Beyond 2 R$_\odot$, the plasmoids were tracked in LASCO/C2 FOV where they travelled at high speeds of 740 to 1800 km s$^{-1}$. In the outer corona the plasmoids showed an average speed of $\sim$1080 km s$^{-1}$. The average speed of upward moving plasmoids show an increase by $\sim$3.5 times when moving from AIA to K-Cor FOV which further increase $\sim$1.5 times on reaching the outher corona. This implies that plasmoids accelerated when propagating outwards. The observed speed range for downward moving plasmoids are $\approx$7-40\% Alfv\'en speed in the inner corona taken as $\sim$1000 km s$^{-1}$ implying their sub-Alfv\'enic nature. On the other hand upward moving plasmoids show speed ranging from 30 to more than 1500 km s$^{-1}$ from inner to outer corona. The wide distribution of speeds for upward moving plasmoids suggests that they become super-Alfv\'enic when propagating outwards. Also the Alfv\'en speed decreases moving farther away from the Sun. This is also an observational evidence supporting the resistive MHD numerical simulation by \citet{Barta2008A&A, Forbes2018ApJ} for the kinematics of plasmoids after magnetic reconnection.

\subsection{Evolution of Plasmoids}

\begin{figure}[!h]
  \centering
  \includegraphics[angle=0,width=9.2cm]{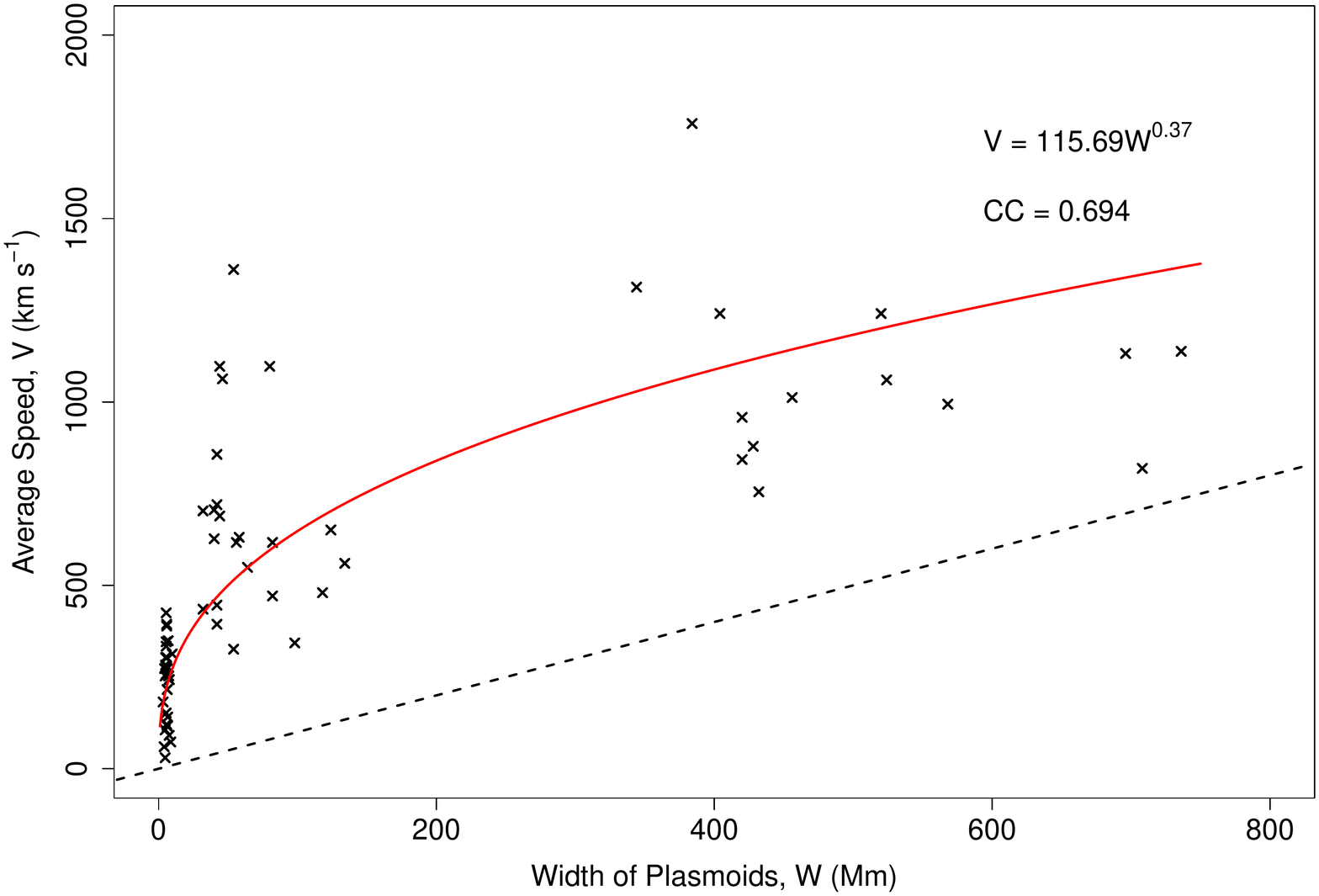}
      \caption{Relation between width and average speed of plasmoids. The red line represents the power law associated with the plasmoid width and average speed. A good correlation with correlation coefficient (CC) of 0.69 is obtained. The dashed line is 1:1 line.}
         \label{fig:cc_speed_size}
  \end{figure}
 
Figure \ref{fig:cc_speed_size} shows the relation between the measured width of the plasmoids with their average speed. The region between size of $\sim$200-400 Mm corresponds to lack of observations in K-Cor FOV to larger heights due to poor SNR. We found that when plasmoids evolve and propagate to outer corona they obey a power-law with the relation,
\begin{equation}
    \label{eq:speed_size}
    V = 115.69W^{0.37},
\end{equation}
where V and W are the average speed and width of the plasmoids. 
The plot shows a scatter with $\sim$69\% correlation between the two quantities. This implies that the average speed of the plasmoids is closely related to their size. However, this relation is valid only to describe the accelerating phase of plasmoid evolution. As the plasmoids intensity reduced in the outer corona, we could not track them further in LASCO/C3 FOV to find out if there is any signature of deceleration later upon propagation. This limits the Equation \ref{eq:speed_size} to explain their evolution characteristic with increasing speeds only.

 \begin{figure}[htbp]
   \centering
   \includegraphics[angle=0,width=9cm]{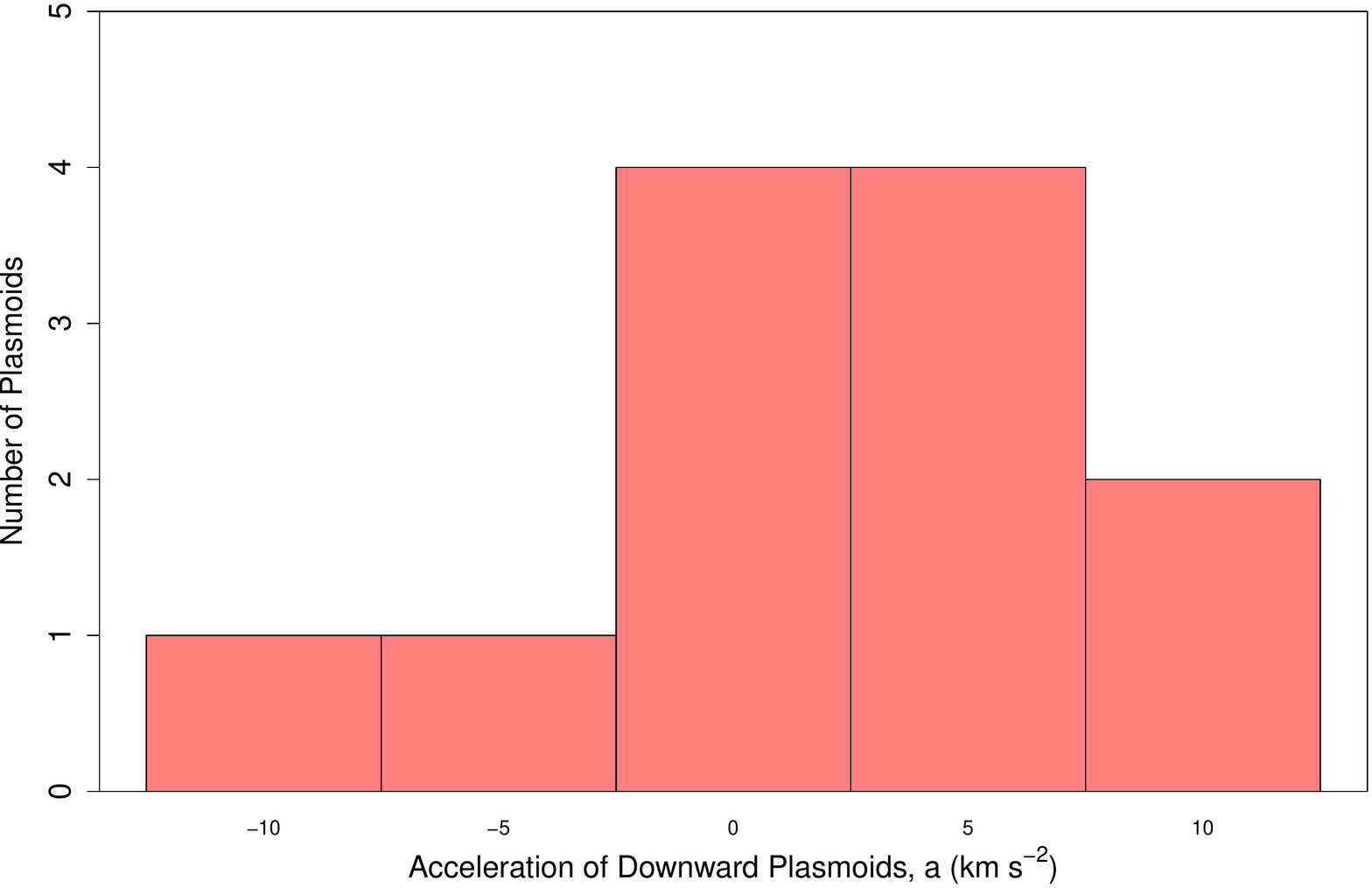}
      \caption{Distribution of the acceleration of the downward moving plasmoids observed in AIA 131 \AA\ images}
         \label{fig:down_acc}
   \end{figure}

In the scenario for the downward moving plasmoids, we also found a variation in speed during their propagation. For 12 out of 20 downward moving plasmoids which could be tracked in atleast three consecutive frames, we could determine their acceleration. It can be seen from Figure \ref{fig:down_acc}  that acceleration of these plasmoids varied from -11 km  s$^{-2}$ to more than 8 km  s$^{-2}$. It was found that a significant fraction (6 out of 12) accelerated on moving downwards, whereas 2 moved with approximately no acceleration. The remaining ones showed deceleration. It was predicted by \citet{Forbes2018ApJ} that the downward moving plasmoids should show acceleration. However, due to the limited observations, this could not be verified in earlier studies \citep{2010cartwheel}. Such observations combined with modelling may add to our knowledge about the true nature of magnetic reconnection.

   

\subsection{Identification of X-point}

 \begin{figure}[htbp]
   \centering
   \includegraphics[angle=0,width=9cm]{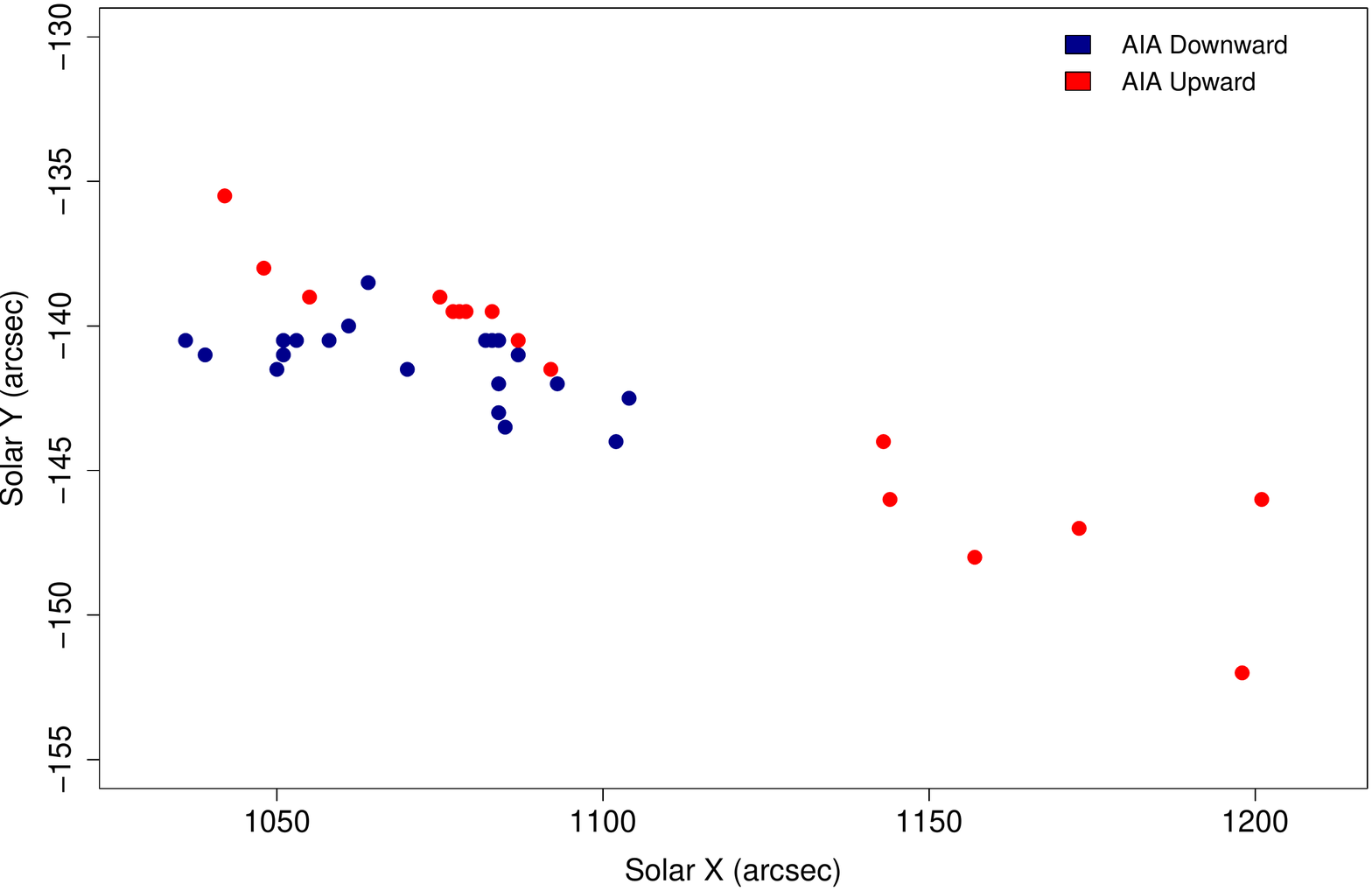}
      \caption{Location of the blobs identified manually from AIA 131 \AA\ passband images. Red corresponds to the position of upward moving plasmoids whereas blue represents the dowward moving ones.}
         \label{fig:blobposi_131}
   \end{figure}
   
We plotted the locations of the first appearance of the upward and downward moving plasmoids in AIA 131 \AA\ images as shown in Figure \ref{fig:blobposi_131}. The points shown in blue marks the initial position of downward moving plasmoids where as red represents upward moving plasmoids. The red points are spread across all the heights along the current sheet in AIA FOV.
The middle panel of Figure \ref{fig: sample_updown} also shows presence of the movements in opposite direction appearing as ridges in the height-time plot.
It should be noted from Figure \ref{fig:blobposi_131} that no downward moving plasmoids appear higher than $\sim$1110 arcsec corresponding to a height of $\approx$0.15 R$_\odot$ from the solar surface. This suggests that the region around 1.15 R$_\odot$ is related to the formation of downward blobs. This region is speculated to be the primary disconnection (X-point) region in the current sheet where reconnection occurred and the flow pairs are generated.
\citet{Forbes2018ApJ} showed that during a reconnection, multiple neutral points may form but a single stagnation point may govern  the dynamics of the current sheet. 
We speculate the stagnation and the primary disconnection point to be located at $\approx$1.15 R$_\odot$.  
It should also be noted that this point is closer to the post-flare loop top which was identified to be around $\sim$1.03 R$_\odot$. This height unlike the models of \citet{Lin2000JGR, Reeves2005ApJ} does not lie at the center of the current sheet length but closer to its lower tip.
{ This estimate is within the range of neutral points identified using height-time plots for the current sheet associated with this event as reported in recent studies \citep{cs_turbulence2018, yu2020magnetic}.}
A null-point height of 1.25 R$_\odot$ was reported in \cite{2010cartwheel} from the identified SADs. A failed attempt to identify a mean null-point height of 1.2 R$_\odot$ was made to explain the origin of post-CME blobs \citep{Schanche2016ApJ}. They extended the possibility of formation of blobs at higher heights than AIA FOV. \citet{Chae2017ApJ} estimated a height of 1.5 R$_\odot$ for the formation of upward moving blobs in their study.

\section{Summary and Discussions}
\label{sec:summary}

Magnetic reconnection plays a major role in small and large scale transients observed in solar atmosphere. In this work we have studied the statistical properties of plasmoids observed in the current sheet at the western solar limb after a fast CME on September 10, 2017. Due to its visibility in multiple pass-bands for a long duration, we have tried to formulate the evolution of plasmoids during reconnection process. In this work we found the Alfv\'en Mach number the ratio of width to length and it turns out to be in the range of 0.018-0.35 which is sufficient to lead an eruption as proposed by \cite{Lin2000JGR} and formation of plasmoids \citep{Vrsnak2003SoPh} during fast reconnection process.
It is observed that the reconnection was accompanied by the plasma blobs along the current sheet in upwards and downward direction in AIA 131 \AA\ pass-band, where as upward moving plasmoids  in white-light coronagraph images of K-Cor and LASCO. 

We have visually identified and tracked these blobs as the bright structures moving along the current sheet in the time series images and developed a statistics. There were 20 downward moving and 16 upwards moving plasma blobs observed in AIA 131 \AA\ images. In the K-Cor images 24 upward moving blobs were observed whereas 17 upward moving plasmoids were identified in LASCO/C2 images. We could not identify any downward moving plasmoids in white-light coronagraph images suggesting that they formed at lower heights corresponding to only AIA FOV.
 { \citet{Lee2020ApJ} reported identification of 34 plasma blobs in K-Cor FOV where as 4 could be identified in LASCO/C2 FOV associated with the current sheet. The difference in the number could be addressed to the manual subjectivity in identification. We would like to mention here that some of the bright propagating bidirectional outflows reported in \citet{yu2020magnetic} could be blobs as these signatures have been identified in our study along with \citet{Lee2020ApJ}. Probably both blobs and flows are produced during the reconnection and co-exists which could be difficult to distinguish at some instances.}

The size distribution of the plasmoids  observed in AIA, K-Cor and LASCO-C2 FOV reveals that it follows a power-law distribution as proposed using resistive-MHD simulation by \citet{guo_plasmoids2013ApJ} and differing from the one found by \cite{McKenzie2011ApJ} for SADs. 
On fitting the power-law to the distribution to the observed widths (W) of plasmoids, we found a decreasing function, $f(W)$ = 218$W^{-1.12}$.
This is the first observational study to support the power-law distribution of plasmoid widths.
This also suggests that the lower limit is determined by the instrument and hence earlier observations could not provide an agreement.
The plasmoids, both upward and downward moving, have average width ranging from $\sim$6 Mm in AIA FOV when they had just formed. These range increases to $\sim$64Mm in white-light K-Cor images which further evolved to $\sim$510 Mm in outer corona in LASCO/C2 FOV. The wide range of sizes hint the tearing of the current sheet over scales from few Mm to hundreds of Mm. These blobs have been identified to be formed in AIA and KCor FOV which evolved to grow in size moving out to the LASCO/C2 FOV. The less number of blobs in LASCO/FOV compared to KCor images also suggests the possibility of merging of blobs to form bigger blobs.
{ The estimated size is larger than the estimations presented in \citet{Lee2020ApJ} corresponding to $\sim$14 to 42 Mm average lateral widths in K-Cor FOV and $\sim$175 Mm in LASCO/C2 FOV. If the average of radial and lateral widths are considered in this study, the estimates of average size becomes $\sim$30 Mm and $\sim$330 Mm for KCor and LASCO/C2 FOV respectively.
The differences could arise due to the differences in the visual identification of the blobs.}

The region separating the upward and downward moving blobs was identified to be located at $\approx$ 1.15 R$_\odot$. This region is regarded as the principal disconnection region where the reconnection is the fastest. \citet{Forbes2018ApJ} suggested that there can be multiple neutral points in a current sheet, but a primary stagnation point will be responsible for the dynamics of the current sheet. Therefore, the height determined in this work can be regarded as this primary stagnation point. As this point lies closer to the post-flare loop top, this also confirms that the primary disconnection and stagnation points lie closer to the lower tip of the current sheet.
Due to difficulty in observation, the search of such region was not successful when plasma blobs identified in LASCO/C2 were tracked back in AIA FOV \citep{Schanche2016ApJ}. Previous study by \citet{2010cartwheel} suggests existence of such region by tracking the downflows of SADs.
For the current sheet formed on September 10, 2017, plasmoids after their formation are found to be moving with an average speed of 272 km s$^{-1}$ towards the Sun and with 191 km s$^{-1}$ in the other direction.
When the upward moving plasmoids had traversed from AIA FOV to $\sim$1.2 to 2 R$_\odot$ and tracked in K-Cor images, the average speed increased to 671 km s$^{-1}$ which further increased to $\sim$ 1080 km s$^{-1}$ beyond 2 R$_\odot$ in LASCO FOV. This suggests that the plasmoids had sufficient energy to get accelerated to super-Alfv\'enic speeds as the reach outer corona. This differs from the earlier result based on limited observation stating that the upward moving plasmoids do not show significant change in speed \citep{2010cartwheel}. 
{ On the other hand a speed range of $\sim$38 to 945 km s$^{-1}$ and 767 to 787 km s$^{-1}$ was measured for the blobs in K-Cor and LASCO/C2 FOV respectively by \citet{Lee2020ApJ}.}
The downward moving plasmoids do not show high speeds as they merge with the underlying loops giving them less time to accelerate unlike the upward moving ones. However, these downward-directed plasmoids show an acceleration ranging from  -11 km  s$^{-2}$ to more than 8 km  s$^{-2}$. The acceleration of the bi-directional plasmoids near the neutral point provides an important implication in determination of the diffusion region in the current sheet \citep{Forbes2018ApJ}.
The speed distribution also satisfies the kinematics properties of upward and downward moving plasmoids proposed by \citet{Barta2008A&A, Forbes2018ApJ}. 
These plasmoids could not be tracked to farther distance due to their decreasing intensity radially outwards in the outer corona. As a result it is still unclear up to what distance after their formation they show acceleration.

It could be seen from Figures \ref{fig:size_all} and \ref{fig:speed_dist_all} that the size of the plasmoids and their speeds are escalated when they have reached outer corona. We tried to relate the two observables and found a good correlation of $\sim$69\% between them. We could establish an empirical relation characterising the two quantities as $V$ = 115.69$W^{0.37}$. The signature of expansion of plasmoids and increase in average speed can be clearly noticed in Figure \ref{fig:cc_speed_size}. This relation based on observational study will be a step forward to constrain the numerical simulation and help us understand the formation and evolution of plasmoids and their role in reconnection process.

The difficulty in tracking the blobs from origin to farther out in the corona was also identified by \citep{Schanche2016ApJ}. The limited availability of white-light coronagraph like K-Cor is also subjected to atmospheric conditions which limits these type of studies. This scenario will be improved in future if similar event is observed by instruments like METIS on board Solar Orbiter \citep{2020A&A...642A...1M}, VELC on Aditya-L1 \citep{ADITYA2017} and ASPIICS of PROBA-3 \citep{Proba3} as they will observe below 3 R$_\odot$.
It is worthwhile to  mention that the limit of white-light observations will be pushed forward with the availability of high resolution images taken at high cadence with good signal to noise ratio by VELC \citep{VELC17} and ASPIICS. It will be helpful to identify the different phases of plasmoids propagation and kinematics properties \citep{Barta2008A&A, Forbes2018ApJ}.
In future, the availability of magnetic field measurements in the inner corona using the VELC payload of Aditya-L1 \citep{Sankarasubramanian2018cosp} will also help us establish the scales of magnetic field in the plasmoids \citep{Huang2012PhRvL}.
This will enable us better understand the plasmoid instability and hence fast reconnection observed in solar atmosphere. 

\begin{acknowledgements}
{ We thank the anonymous referee for the valuable comments.} VP is supported by the Spanish Ministerio de Ciencia, Innovaci\'on y Universidades through project PGC2018-102108-B-I00 and FEDER funds. VP was further supported by the GOA-2015-014 (KU Leuven) and the European Research Council (ERC) under the European Union's Horizon 2020 research and innovation programme (grant agreement No 724326).
V.P. acknowledge support from the International Space Science Institute (ISSI), Bern, Switzerland to the International Team 413 ``Large-Amplitude Oscillations as a Probe of Quiescent and Erupting Solar Prominences'' (P.I. M. Luna) which enabled thoughtful discussions with Dr. Judith Karpen.
K.C. is supported by the Research Council of Norway through its Centres of Excellence scheme (project number 262622).
We also thank Dr. Daniel B. Seaton, Bibhuti Kumar Jha, and Satabdwa Majumdar for the discussions and suggestions.
We acknowledge NASA/SDO team to make AIA data for open access. SDO is a mission for NASA's Living With a Star (LWS) program. SOHO is a project of international collaboration between ESA and NASA. We also thank Mauna Loa Solar Observatory located at High Altitude Observatory for making K-Cor data available.
     
\end{acknowledgements}

\bibliographystyle{aa} 
\bibliography{bib_cs} 
\end{document}